\documentclass[aps,pre,twocolumn,showpacs,unsortedaddress,floatfix]{revtex4-1} 
\usepackage{graphicx}

\begin{document}
\title{Collective effects at frictional interfaces}
\author{O.M. Braun}
\email[E-mail: ]{obraun.gm@gmail.com}
\homepage[Web: ]{http://www.iop.kiev.ua/~obraun}
\affiliation{Institute of Physics, National Academy of Sciences of Ukraine,
  46 Science Avenue, 03028 Kiev, Ukraine}
\author{Michel Peyrard}
\email[E-mail: ]{Michel.Peyrard@ens-lyon.fr}
\affiliation{Laboratoire de Physique de l'Ecole Normale Sup\'{e}rieure de Lyon,
  46 All\'{e}e d'Italie, 69364 Lyon C\'{e}dex 07, France}
\author{D.V. Stryzheus}
\affiliation{Institute of Physics, National Academy of Sciences of Ukraine,
  46 Science Avenue, 03028 Kiev, Ukraine}
\author{Erio Tosatti}
\affiliation{International School for Advanced Studies (SISSA),
  Via Bonomea 265, I-34136 Trieste, Italy}
\affiliation{International Centre for Theoretical Physics (ICTP),
  P.O. Box 586, I-34014 Trieste, Italy}
\affiliation{CNR-IOM Democritos National Simulation Center,
  Via Bonomea 265, I-34136 Trieste, Italy}
\begin{abstract}
We discuss the role of the long-range elastic interaction between the
contacts inside an inhomogeneous frictional interface.  The interaction
produces a characteristic elastic correlation length $\lambda_c = a^2 E
/ k_c$ (where $a$ is the distance between the contacts, $k_c$ is the
elastic constant of a contact, and $E$ is the Young modulus of the
sliding body), below which the slider may be considered as a rigid body.
The strong inter-contact interaction leads to a narrowing of the
effective threshold distribution for contact breaking and enhances the
chances for an elastic instability to appear.  Above the correlation
length, $r > \lambda_c$, the interaction leads to screening of local
perturbations in the interface, or to appearance of collective modes ---
frictional cracks propagating as solitary waves.


\end{abstract}
% 46.55.+d (Tribology and mechanical contacts)
% 81.40.Pq (Friction, lubrication and wear in materials science)
% 61.72.Hh (indirect dislocations,slip,creep,friction)

\maketitle

%=========================================================================
\section{Introduction}
\label{intro}

Studies of sliding friction,
a subject with great practical importance
and with rich physics,  attracted an increased
interest during last two decades~\cite{P0,BN2006}.
Tip-based experimental techniques as well as
atomistic molecular dynamics (MD) computer simulations
describe with considerable success
the processes and mechanisms
operating in atomic-scale friction.
Much less is known when going to meso- and macro-scale % sliding
friction, where one has to take into account that
the frictional interface is inhomogeneous and generally complex.
An immediate example is dry friction between rough surfaces.
Even when the sliding surfaces are ideally flat but for example
the substrates are not monocrystalline,
or there is an interposed solid lubricant film consisting of 
misoriented domains,
the frictional interface is again inhomogeneous.
The same may be true even for liquid lubrication,
if under applied load
the lubricant solidifies making bridges due to
Lifshitz-Sl\"ozov coalescence.
In these cases, the so-called
earthquake-like (EQ) type models
can be successfully applied~\cite{BN2006,
OFC1992,P1995,BR2002,FKU2004,FDW2005,BP2008,BT2009,BBU2009,BP2010,%
BP2011,BT2011}. 
In the EQ model, the two (top and bottom) mutually sliding surfaces
are coupled by a set of contacts, representing, e.g., asperities, patches
of lubricant, or 2D crystalline domains.
A contact is assumed to behave
as a spring of elastic constant $k_c$ so long as
its length is shorter than a critical value $x_s = f_s /k_c$;
above this length the contact breaks, to be subsequently restored
with % zero length and
lower stress.
The sliding kinetics of this model may be reduced to a master equation (ME),
which allows an analytical study~\cite{BP2008,BP2010,BP2011}.

In the simplest approach, the slider is treated as a rigid body.
Due to the non-rigidity of the substrates, however,
several length scales naturally appear  in the problem.
First, different regions of the interface will exhibit different displacements.
The length $\lambda_{L}$ such that for distances
$r \gg \lambda_L$ the displacements are independent,
is known as the Larkin-Ovchinnikov length~\cite{LO1979}.
It was shown~\cite{PT1999} that for the contact of stiff rough solid surfaces
$\lambda_L$ may reach unphysically large values
$\sim 10^{100\,000}$~m.
Second, deformation of the solid substrates
leads to the elastic interaction between the contacts.
Elasticity will correlate variations of forces on nearest contacts
over some length $\lambda_c$ known as the elastic correlation
length~\cite{CN1998}. 
Third, displacements in one region of the slider will be felt
in other regions on the distance scale set by of
a screening length $\lambda_s$.
Finally, the breaking of one contact may stimulate neighboring contacts
to break too (the so-called concerted, or cascade jumps),
following which
an avalanche-like collective motion of different domains of the interface
may appear~\cite{BR2002}.

In this paper we discuss collective effects in the frictional interface
and propose approaches to treat them from different viewpoints.
In particular, our aim is to clarify the following questions:
(\textit{i})~what is the law of interaction between the contacts;
(\textit{ii})~at which scale can the slider be considered as a rigid body, or
what is the coherence distance $\lambda_c$ within which
the motion of contacts is strongly correlated;
(\textit{iii})~whether  the interaction effects
can be incorporated in the master equation approach;
(\textit{iv})~how does the interaction modifies the interface dynamics;
(\textit{v})~what is the screening length $\lambda_s$;
(\textit{vi})~when do avalanche motion of contacts (a self-healing crack) 
appear, and what is the avalanche velocity?

The paper is organized as follows.
The earthquake-like model, its description with the ME approach,
and the elastic instability responsible for the stick-slip motion
are introduced in Sec.~\ref{EQ}.
The interaction between contacts is studied in Sec.~\ref{interact}.
An approach to incorporate the interaction between contacts into the ME approach
in a mean-field fashion is described in Sec.~\ref{MF}.
The role of interaction at the meso/macro-scale is considered in
Sec.~\ref{farzone}. 
Finally, discussions in Sec.~\ref{concl} conclude the paper.

%=========================================================================
\section{Earthquake-like model, master equation and elastic instability}
\label{EQ}

%-------------------------------------------------------------------------
\subsection{The earthquake-like model}

In the EQ model
the sliding interface is treated as a set of $N$ contacts
which deform elastically with the average rigidity $k_c$.
The $i$th contact connects the slider and the substrate through
a spring of shear elastic constant $k_i$.
When the slider is moved, the
position of each contact point changes, the contact spring elongates (or
shortens) so that the slider experiences a force
$-F = \sum f_i$  from the interface,
where $f_i = k_i x_{i}$ and $x_{i} (t)$
is the shift of the $i$th junction from its unstressed position.
The contacts are assumed to be coupled ``frictionally'' to the slider.
As long as the force $|f_i|$ is below a certain threshold $f_{si}$,
the $i$th contact moves together with the slider.
When the force exceeds the threshold, the contact breaks and
a rapid local slip takes place, during which the local stress drops.
Subsequently the junction is pinned again
in a less-stressed state with $f_{bi}$, and the whole process repeats itself.
Thus, with every contact we associate the threshold value $f_{si}$
and the backward value $f_{bi}$,
which take random values from the distributions
$\widetilde{P}_c (f)$ and $\widetilde{R} (f)$ correspondingly.
When a contact is formed again (re-attached to the slider),
new values for its parameters are assigned.
The EQ model was studied numerically in a number of
works~\cite{% BK1967,CL1989,
OFC1992,P1995,BR2002,FKU2004,FDW2005,BP2008,BT2009,BBU2009},
typically with the help of % some
the cellular automaton numerical algorithm.

%-------------------------------------------------------------------------
\subsection{The master equation approach}

Rather than studying the evolution of the EQ model by numerical simulation,
it is possible to describe it analytically~\cite{BP2008,BP2010,BP2011}.
Let $P_c (x)$ be the normalized probability distribution of values
of the stretching thresholds $x_{si}$ at which contacts break;
it is coupled with the distribution of threshold forces by the relationship
$P_c (x) \, dx = \widetilde{P}_c (f) \, df$, i.e.,
the corresponding distributions are coupled by the relationship
$P_c (x) \propto x \, \widetilde{P}_c [f(x)]$,
where $f \propto x^2$~\cite{BP2010}.
We assume that the distribution $P_c (x)$ has a dispersion $\Delta x_s$
centered at $x=x_c$.

To describe the evolution of the model,
we introduce the distribution $Q(x;X)$ of the contact stretchings $x_i$
when the % bottom of the
sliding block is at position $X$.
Evolution of the system is described
by the integro-differential equation
(known as the master equation, or the kinetic equation,
or the Boltzmann equation)~\cite{BP2008,BP2010}
\begin{equation}
\left[ \frac{\partial}{\partial X} +
\frac{\partial}{\partial x}
+ P(x) \right] Q(x; X) = R(x) \, \Gamma (X) \,,
\label{mf.eq02}
\end{equation}
where
\begin{equation}
\Gamma (X) = \int_{-\infty}^{\infty} d\xi \, P(\xi) \, Q(\xi; X)
\label{mf.eq03}
\end{equation}
and
\begin{equation}
P(x) = P_c (x) / J_c (x) \,,
\;\;\;
J_c (x) = \int_x^{\infty} d\xi \, P_c (\xi) \,.
\label{mf.eq04}
\end{equation}
Then, the friction force
(the total force experiences by the slider from the interface)
is given by ($k_c = \langle k_i \rangle$)
\begin{equation}
F(X)=N k_c \int_{-\infty}^{\infty} dx \, x \, Q(x;X) \,.
\label{mf.eq05}
\end{equation}

In the steady state corresponding to smooth sliding, the ME reduces to
\begin{equation}
d Q(x) / dx + P(x) \, Q(x) = R(x) \, \Gamma \,,
\label{mf.eq06}
\end{equation}
which has the solution
\begin{equation}
Q_s (x) = {\cal N}
E_P (x) \left[
1 + \Gamma \int_{0^+}^{x}
d\xi \, R(\xi) / E_P (\xi)
\right] ,
\label{mf.eq07}
\end{equation}
where ${\cal N}$ is the normalization constant,
$\int_0^{\infty} dx \, Q_s (x) =1$,
and
\begin{equation}
E_P (x) = \exp \left[ -U(x) \right] ,
\;\;\;
U(x) = \int_{0}^{x} d\xi \, P(\xi) \,.
\label{mf.eq08}
\end{equation}

%-------------------------------------------------------------------------
\subsection{Elastic instability}
\label{instability}

The solution of the ME ~\cite{BP2008,BP2010} shows that
when a {\it rigid\/} slider
begins to move adiabatically, $\dot{X} >0$,
it experiences from the interface a friction force $F_{\infty}(X) <0$.
% typically shown as a function of position in Fig.~\ref{A03-Erio}.
Initially $|F_{\infty}|$ grows roughly linearly with $X$,
$|F_{\infty}| \approx K_s X$
(here $K_s = N k_c$ is the total % force
elastic constant (``rigidity'') of the interface),
until it reaches a value $\sim F_s - \Delta F_s$,
where $F_s \approx K_s x_c$ and $\Delta F_s \approx K_s \Delta x_s$.
Gradually however contacts begin to break and reform, slowing down the
increase of $|F_{\infty}|$ and then inverting the slope through a displacement
$\Delta x_s$ until almost all 
contacts have been reborn. Successively the process repeats itself with
a smaller 
amplitude until, due to increasing dispersion of breaking and reforming
processes, 
the force asymptotically levels off and attains
a position independent steady state kinetic friction value % $F_k$
with smooth sliding.

According to Newton's third law,
the external driving force $F_d  = K (vt-X)$ which causes the displacement $X$
(here $K$ is the slider rigidity and $v$ is the driving velocity),
is compensated
by the force from the interface, $F_d = F(X)$.
Smooth sliding
is always attained with a rigid slider.
It persists for a nonrigid slider as well,
so long as the pulling spring
stiffness is large enough, $K > K^*$, where
\begin{equation}
K^* = \, {\max} \, F^{\prime}_{\infty} (X) \,,
\;\;\;
F^{\prime}_{\infty} (X) \equiv d F_{\infty} (X)/dX \,.
\label{mf.eq13}
\end{equation}
When conversely the slider, or the pulling spring elastic constant
are soft enough  ($K < K^*$)
there is a mechanical instability.
The driving force $F_d$
cannot be compensated by the force from the interface, and
the slider motion becomes unstable at $X_c$, where $X_c$ is the (lowest)
solution of $F^{\prime}_{\infty}(X) = K$
(for details see Refs.~\cite{BP2008,BP2010}).
The mechanical instability yields stick-slip frictional motion of the slider.

Thus, the regime of motion --- either
stick-slip for $K \ll K^*$ or smooth sliding for $K \gg K^*$
--- is controlled by the effective stiffness parameter
% \begin{equation} \label{}
$K^* \sim K_s x_c / \Delta x_s$.
% \end{equation}
When all contacts are identical, $\Delta x_s =0$ so that
$K^* =\infty$, one always obtains a stick-slip motion.

%-------------------------------------------------------------------------
\subsection{Material parameters}
\label{material}

It is % seems
useful here, before proceeding with the analytical and numerical developments
necessary to answer the questions posed in the Introduction,
to review the practical significance and magnitude of
the % different and numerous parameters of the model.
model parameters.

\smallskip
\textit{Elastic constant of the slider}.
% When the driving force is applied to the top of the slider,
The slider (shear) elastic constant $K$ is equal to
% \begin{equation} \label{K1}
$K % \equiv {F}/{\Delta X}
= [{E}/{2 \, (1+\sigma)}] [{L_x L_y}/{H}]$,
% \end{equation}
where
$L_x$, $L_y$ and $H$ are the slider dimensions,
$E$ and $\sigma$ are the substrate Young modulus and
Poisson ratio, respectively~\cite{LL1986}.
For example, for a steel slider of Young's modulus
$E=2 \times 10^{11}$~N/m$^2$,
Poisson's ratio $\sigma =0.3$
and the size
$L_x \times L_y \times H = 1\;$cm$\;\times \; 1\;$cm$\;\times \;1\;$cm,
we obtain $K \sim 10^{9}$~N/m.

\smallskip
\textit{Rigidity of the interface contacts}.
Here we characterize the typical magnitudes
of the contact stretching length $x_c$ and stiffness $k_c$.
Assume the slider and the substrate to be coupled by
$N = L_x L_y /a^2$ contacts,
and that the contacts have a cylindrical shape of (average) radius $r_c$
with a distance $a$ between the contacts.
It is useful to introduce the dimensionless  parameter $\gamma_2 = r_c /a$,
% ($\gamma_2 < 0.5$)
which may be estimated as follows~\cite{P0}.
Consider a cube of linear size $L$ on a table.
The weight of the cube $F_l = \rho L^3 g$
($\rho$ is the mass density and $g=9.8$~m/s$^2$)
must be compensated by forces from the contacts, $F_l = N r_c^2 \sigma_c$,
where $\sigma_c$ is the plastic yield stress. Then,
$\gamma_2^2 = (N r_c^2)/(N a^2) = (\rho L^3 g)/(\sigma_c L^2)$, or
% \begin{equation} \label{eq14}
$\gamma_2 = (\rho L g/ \sigma_c )^{1/2}$.
% \end{equation}
Taking $L = 1$~cm, $\rho = 10$~g/cm$^3$ and
$\sigma_{c} = 10^9$~N/m$^2$ (steel),
we obtain $\gamma_{2} \approx 10^{-3}$
which should be typical for a contact of rough stiff surfaces.
For softer materials, and especially for a lubricated interface,
the values of $\gamma_2$ would be much larger, e.g., $\gamma_2 \sim 0.1$.

The second dimensionless parameter
$\gamma_1 = k_c / E a$
characterizes the stiffness of the contacts.
To estimate $\gamma_1$, assume again contacts
with the shape of a cylinder of radius
$r_c$ and length $h$ ($h$ is the thickness of the interface).
Suppose in addition that one end of a contact (``column'') is fixed,
and a shear force $f$ is applied to the free end.
This force will lead to the displacement $x = f/k_c$ of the end,
where $k_c = 3E_c I/h^3$, $E_c$ is the Young modulus of the contact material
and $I = \pi r_c^4 /4$ is the moment of inertia of the cylinder~\cite{LL1986}.
% (see Ref.~\cite{LL1986}).
% , Sec.~II.17, Eq.~(17.11), p.~97 and Sec.~II.20, exercise~3, p.~116
In this way we obtain
% \begin{equation} \label{eq15}
$k_c = (3\pi /4)(E_c r_c)(r_c /h)^3$,
% \end{equation}
so that
% \begin{equation} \label{eq13}
$\gamma_1 = (3\pi /4) \left( E_c a^3 / E h^3 \right)
\left( r_c / a \right)^4 = \gamma_0 \gamma_2^4$ with
$\gamma_0 = (3\pi /4)(E_c /E)(a /h)^3$.
For the contact of rough surfaces,
where $E_c =E$ and $a \agt h$, we have $\gamma_0 \agt 1$,
while for lubricated interfaces where $E_c \ll E$,
one would expect $\gamma_0 \alt 1$.

An estimate of characteristic values~\cite{P0} leads to
$r_c \sim (10^{-3} \div 10^{-2}) \, a$.
Thus, for the steel slider considered above, taking
$r_c = h = 1$~$\mu$m and intercontact spacing
$a = 3 \times 10^2 \, r_c$, we obtain
$N \sim 10^3$ and
$k_c \sim 5 \times 10^5$~N/m, so that
the global stiffness of the interface is
$K_s  \sim 5 \times 10^8$~N/m.

\smallskip
\textit{Stick-slip versus smooth sliding}.
As mentioned above in Sec.~\ref{instability},
the regime of motion (either stick-slip or smooth sliding)
is controlled by the % critical stiffness
parameter
$K^* \sim K_s x_c / \Delta x_s$.
For the steel slider considered above, estimates gave
$K \sim 10^9$~N/m and $K_s \sim 5 \times 10^8$~N/m.
Thus, if the surfaces are rough so that $\Delta x_s \sim x_c$,
then $K  >  K^*$ and one should typically get smooth sliding.
Stick slip appears further disfavored if we consider
a realistic $P_c (x)$ distribution.
For all cases mentioned in Introduction ---
the contact of rough surfaces (both for elastic or plastic asperities),
the contact of polycrystal (flat) substrates, and
the case of lubricated interface,
when the lubricant, melted during a slip,
solidifies and forms bridges at stick,
% forming bridges due to growing of solid grains of the lubricant
--- the distribution $P_c (x)$ is rather wide
with a large concentration of small-threshold contacts~\cite{BP2010},
which makes the value of $K^*$ very small.
Thus, the theory predicts that most systems do not
undergo an elastic instability
and should not therefore exhibit stick-slip.
This conclusion contradicts everyday experience
as well as careful experiments, where stick-slip is pervasive.
As suggested by EQ simulations~\cite{BT2011}, the discrepancy
is most likely caused by ignoring
the elastic interaction between the contacts.

The role of interaction is considered in the next sections.
First, however, we
need to define the form and parameters of the interaction between contacts.

%=========================================================================
\section{Interaction between contacts}
\label{interact}

\begin{figure} [h] %[t] \bigskip
\includegraphics[clip,width=2.8cm,height=4.7cm]{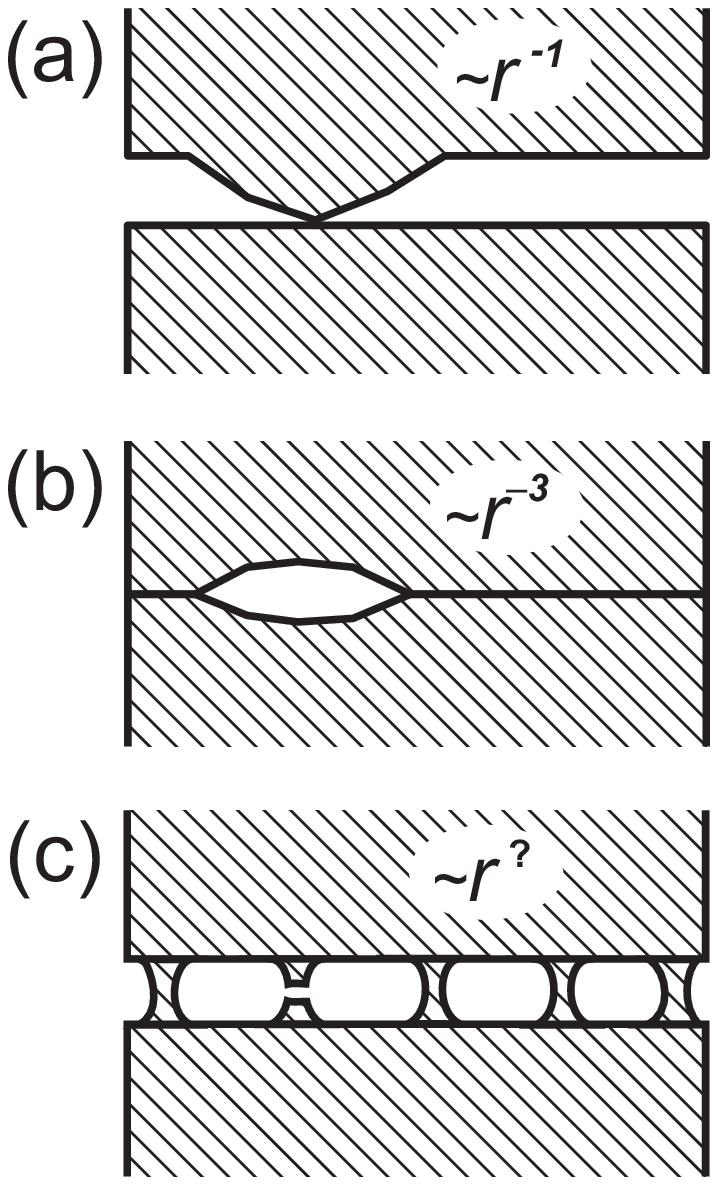} % clip
\hspace{0.3cm}
\includegraphics[clip,width=4.7cm]{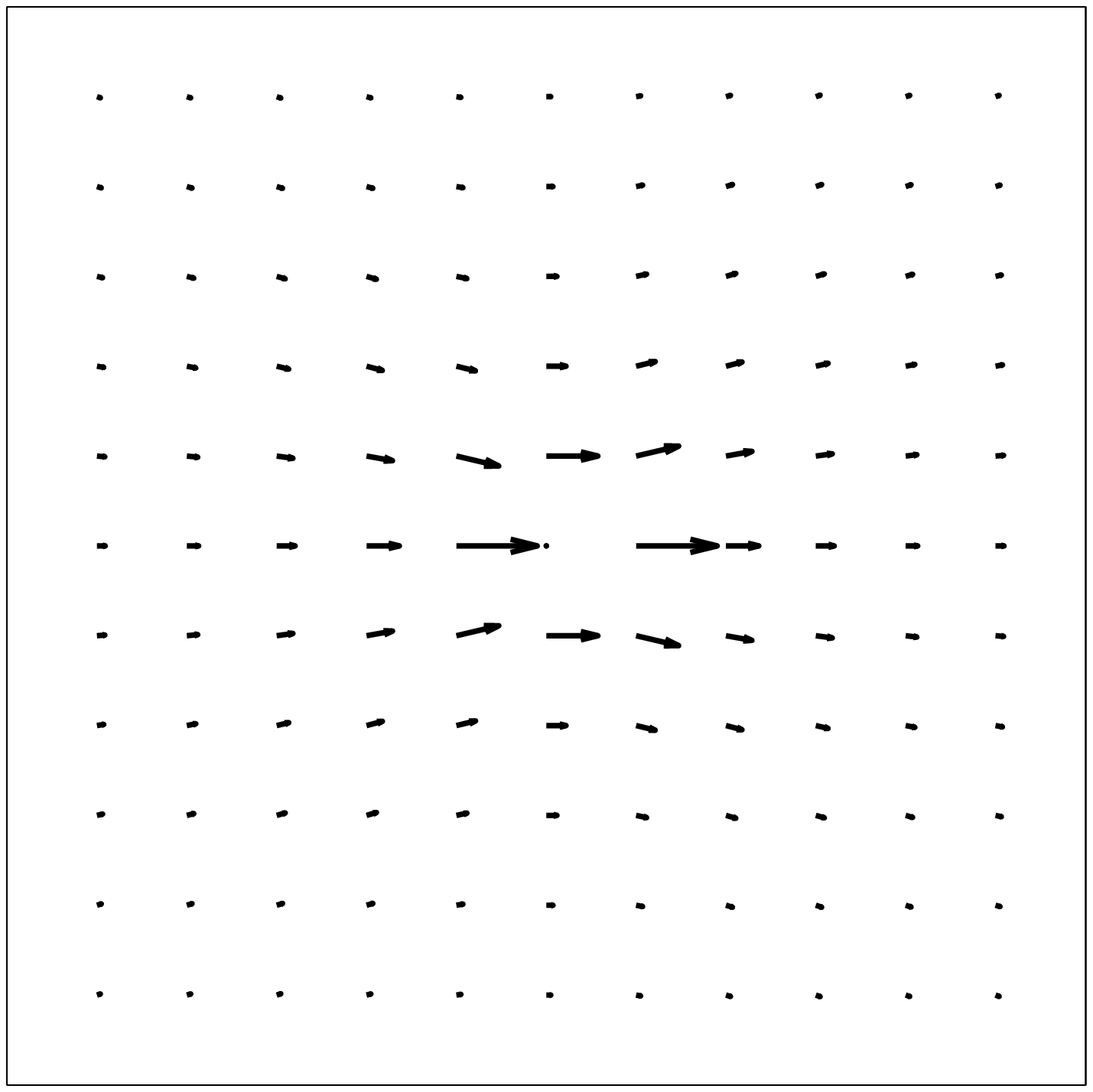}
\caption{\label{int.A10}
Left: decaying of the displacement field at the interface (schematic):
(a) for a single contact $u(r) \propto r^{-1}$,
(b) for a single hole $u(r) \propto r^{-3}$, and
(c) for the array of contacts.
Right: change of forces on contacts when the central contact is removed
($\gamma_1 = 0.06$).
%[use int.A01, int.A10]
}
\end{figure}
Friction is not a simple sum of individual contact
properties. The collective behavior of the contacts is important.
Recently Persson~\cite{P2001a,PBC2002,P2008,YP2008,ACPP2011} developed a
contact mechanics theory
based on the fractal structure of surfaces in order to determine
the actual contact area at all length scales, which determines the
friction coefficient. This approach
includes the presence of multiple contacts and leads to the correct
low-threshold limit $\widetilde{P}_c (f \to 0) =0$.
Persson found that the distribution
of normal stresses $\sigma$ ($\sigma >0$) at the interface
may approximately be described by the expression
$P_{\sigma} ({\sigma}) \propto
\exp \left[ - (\sigma - \bar{\sigma})^2/\Delta \sigma^2 \right] -
\exp \left[ - (\sigma + \bar{\sigma})^2/\Delta \sigma^2 \right]$,
where $\bar{\sigma}$ is the nominal squeezing pressure,
the distribution width is given by $\Delta \sigma = E^* {\cal R}^{1/2}$
($E^*$ is the combined Young modulus of the substrates,
$E^{*-1}=E_1^{-1}+E_2^{-1}$ where $E_{1,2}$ are the Young modula
of the two substrates),
and the parameter ${\cal R}$ is determined by the roughness of the
contacting surfaces,
${\cal R} = (4\pi)^{-1} \int dq \; q^3 \int d^2 x \; \langle h({\mathbf x})
h({\mathbf 0}) \rangle e^{-i {\mathbf q} {\mathbf x}}$.
Assuming that a local shear threshold is directly proportional to the
local normal stress, $f \propto \sigma$, we finally obtain
the distribution, which is characterized by a low concentration
of small shear thresholds,
$\widetilde{P}_c (f) \propto f$ at $f \to 0$,
and a fast decaying tail,
$\widetilde{P}_c (f) \propto \exp (-f^2/f^{*2})$ at $f \to \infty$,
i.e., now the peaked structure of the distribution
is much more pronounced.

\medskip
However an important aspect which has to be included is the
redistribution of the forces when some contacts deform or break.
A concerted motion of contacts may emerge
only due to interaction between the contacts which occurs through the
deformation of the bulk in the directions parallel to the average
contact plane. It is this aspect that we want to consider here.
For the elastic interaction, a qualitative picture is presented
in Fig.~\ref{int.A10} (left).
When a contact acts on the surface at $r=0$ with a force $f$,
it produces a displacement field $u(r) \propto r^{-1}$
which affects other contacts (Fig.~\ref{int.A10}a)
--- similar to the
Coulomb potential for a point charge~\cite{LL1986}.
However, if there are two surfaces, then the same contact acts on the second
surface with the opposite force $-f$ and, if the two surfaces are in contact,
the resulting displacement field should fall as $u(r) \propto r^{-3}$
(Fig.~\ref{int.A10}b)
--- similar to the dipole-dipole potential for a screened point charge
near a metal surface~\cite{PALB2004}.
The question thus is the form of the interaction for the
multi-contact interface (Fig.~\ref{int.A10}c).
We will show that the interaction between the contacts
has a crossover from the~$r^{-1}$
slow Coulomb decay at short distances to the faster dipole-dipole one
at large distances. % $r \to \infty$.

%-------------------------------------------------------------------------
\subsection{Analytics}
\label{int.analytics}

Let us consider an array of $N$ elastic contacts (springs)
with coordinates $\mathbf{r}_i \equiv \{ x_i, y_i, 0 \}$, $i=1,\ldots,N$,
between the two (top and bottom) substrates.
If the interface is in a stressed state,
the contacts act on the top substrate with forces
$\mathbf{f}_i \equiv \{ f_{ix}, f_{iy}, f_{iz} \}$.
These contact forces produce displacements $\mathbf{u}_i^{\rm (top)}$
of the (bottom) surface of the top substrate.
The $3N$-dimensional vectors
$\mathbf{U}^{\rm (top)} \equiv \{ \mathbf{u}_i^{\rm (top)} \}$ and
$\mathbf{F}_t \equiv \{ \mathbf{f}_i \}$
are coupled by the linear relationship
% \begin{equation} \label{int.eq01}
$\mathbf{U}^{\rm (top)} = \mathbf{G}^{\rm (top)} \mathbf{F}_t$.
% \end{equation}
Elements of the elastic matrix $\mathbf{G}^{\rm (top)}$
(known also as the elastic Green tensor)
for a semi-infinite isotropic substrate were given
by Landau and Lifshitz~\cite{LL1986}: % (Sec.~I.8):
\begin{eqnarray}
\label{int.eq02}
\begin{array}{l}
   G_{ix,jx} = g(r_{ij}) [ 2 (1-\sigma) + 2 \sigma x_{ij}^2 /r_{ij}^2 ]
\\ G_{ix,jy} = 2 g(r_{ij}) \, \sigma x_{ij} y_{ij} /r_{ij}^2
\\ G_{ix,jz} = - g(r_{ij}) (1-2\sigma) \, x_{ij} / r_{ij}
\\ G_{iz,jx} = -G_{ix,jz}
\\ G_{iz,jz} = 2 g(r_{ij}) (1-\sigma) \,,
% \\ \ldots \,\,\, ,
\end{array}
\end{eqnarray}
where $x_{ij} = x_i - x_j$, $g(r) = (1+\sigma)/(2 \pi E r)$, and
$\sigma$ and $E$ are the Poisson ratio and Young modulus
of the top substrate, respectively.

In the equilibrium state,
the forces that act from the contacts on the bottom substrate,
must be equal to $\mathbf{F}_b = -\mathbf{F}_t$
according to Newton's third law.
These forces lead to displacements
of the (top) surface of the bottom substrate,
$\mathbf{U}^{\rm (bottom)} = -\mathbf{G}^{\rm (bottom)} \mathbf{F}_t$.
The elements of the bottom Green tensor $\mathbf{G}^{\rm (bottom)}$
are defined by the same expressions (\ref{int.eq02})
except the $xz$ elements for which
$G_{ix,jz}^{\rm (bottom)} = -G_{ix,jz}^{\rm (top)}$
(if the substrates are identical, the $z$ displacements are irrelevant).
Thus, the relative displacements at the interface
due to elastic interaction between the contacts
are determined by %the relation
\begin{equation}
\label{int.eq04}
\mathbf{U} \equiv \mathbf{U}^{\rm (top)} - \mathbf{U}^{\rm (bottom)}
= - \mathbf{G} \mathbf{F} \,,
\end{equation}
where $\mathbf{F} = -\mathbf{F}_t$ and
% \begin{equation} \label{int.eq05}
$\mathbf{G} = \mathbf{G}^{\rm (top)} + \mathbf{G}^{\rm (bottom)}$.
% \end{equation}

On the other hand, the forces and displacements are coupled
by the diagonal matrix (the contacts' elastic matrix) $\mathbf{K}$,
$K_{i \alpha, \, j \beta} = k_{i  \alpha} \delta_{i j} \delta_{\alpha \beta}$
($\alpha, \beta = x,y,z$):
\begin{equation}
\label{int.eq06}
\mathbf{F} = \mathbf{K} \, ( \mathbf{U}_0 + \mathbf{U} ) \,,
\end{equation}
where $\mathbf{U}_0$ defines a given stressed state
(because of linearity of the elastic response,
final results should not depend of $\mathbf{U}_0$).
The total force at the interface,
$\mathbf{f} = \sum_i \mathbf{f}_i$,
must be compensated by external forces applied to the substrates,
e.g., by the force $\mathbf{f}^{\rm (ext)} = \mathbf{f}$ applied
to the top surface of the top substrate
if the bottom surface of the bottom substrate is fixed.

Combining Eqs.~(\ref{int.eq04}) and~(\ref{int.eq06}), we obtain
$\mathbf{F} = \mathbf{K} \, ( \mathbf{U}_0 - \mathbf{G} \mathbf{F} )$,
or
\begin{equation}
\label{int.eq08}
\mathbf{F} = \mathbf{B} \mathbf{K} \mathbf{U}_0 \,,
\;\;\;{\rm where}\;\;\;
\mathbf{B} = ( \mathbf{1} + \mathbf{K} \mathbf{G} )^{-1}.
\end{equation}

If one changes the contact elastic matrix,
$\mathbf{K} \to \mathbf{K} + \mathbf{\delta K}$,
then the interface forces should change as well,
$\mathbf{F} \to \mathbf{F} + \mathbf{\delta F}$.
From Eq.~(\ref{int.eq08}) we have
$ \mathbf{\delta F} = (\mathbf{\delta B}) \mathbf{K} \mathbf{U}_0 +
\mathbf{B} (\mathbf{\delta K}) \mathbf{U}_0 $.
Then, $\mathbf{\delta B}$ may be found from the equation
$ \delta [ \mathbf{B} \, ( \mathbf{1} + \mathbf{K} \mathbf{G} ) ] =
(\mathbf{\delta B} ) ( \mathbf{1} + \mathbf{K} \mathbf{G} ) +
\mathbf{B} (\mathbf{\delta K}) \mathbf{G} = 0 $.
Therefore, finally we obtain:
\begin{equation}
\label{int.eq09}
\mathbf{\delta F} =
\mathbf{B} \, \mathbf{\delta K} \,
(\mathbf{1} - \mathbf{G} \mathbf{B} \mathbf{K}) \mathbf{U}_0 \,.
\end{equation}
Above we have assumed that $\mathbf{\delta K}$ is small.
If it is not small, we have to use the expression
% \begin{equation} \label{int.eq09a1}
$\mathbf{\delta F} =
\mathbf{B} \, \mathbf{\delta K} \,
(\mathbf{1} - \mathbf{G} \mathbf{B} \widetilde{\mathbf{K}}) \mathbf{U}_0$,
% \end{equation}
where
% \begin{equation} \label{int.eq09a2}
$\widetilde{\mathbf{K}} =
\left( \mathbf{1} + \mathbf{\delta K G B} \right)^{-1}
\left( \mathbf{K} + \mathbf{\delta K} \right)$.
% \end{equation}

Now, if we remove the $i^*$th contact by putting
$\delta k_{i \alpha} = - k_{i \alpha} \delta_{i i^*}$
and then calculate the resulting change of forces on other contacts,
we can find a response of the interface to the breaking of a single contact
as a function of the distance
$\mathbf{r} = \mathbf{r}_i - \mathbf{r}_{i^*}$
from the broken contact.

%-------------------------------------------------------------------------
\subsection{Numerics}
\label{int.numerics}

Equation (\ref{int.eq09}) may be solved numerically
by standard methods of matrix algebra.
We explore an idealized array of identical contacts,
$k_{i \alpha} = k_c$ and
$(\mathbf{U}_0)_{i \alpha} = u_0 \, \delta_{\alpha x}$ for all $i$,
organized in a square $89 \times 89$ lattice
with spacing $a=1$,
with the broken contact $i^*$ at the center of the lattice.
For singular terms of the Green function~(\ref{int.eq02})
we apply a cutoff at $r_{ii} = r_c$.
Numerical results depend on two dimensionless parameters.
The first is
$\gamma_1 = k_c / E_* a$, which
determines the stiffness of the array of contacts
relative the substrates
(here $E_*^{-1} = E_{\rm top}^{-1} + E_{\rm bottom}^{-1}$).
The second parameter
$\gamma_2 = r_c /a$
characterizes a single contact
(or the density of asperities).
For the Poisson ratio we took a typical value $\sigma =0.3$.
A typical distribution of breaking induced force changes
is shown in Fig.~\ref{int.A10} (right).

\begin{figure} [h] %[t] \bigskip
\includegraphics[clip,width=8cm]{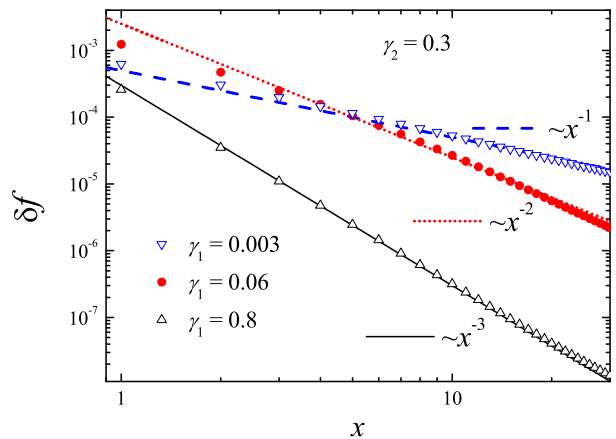}
\caption{(Color online):
\label{int.A02}
Dependence of the change of forces $\delta \! f (r)$
on the distance $x$ from the broken contact
for three values of the interface stiffness:
$\gamma_1 =0.003$ (blue down triangles, dashed line),
0.06 (red solid circles, dotted line) and 0.8 (black up triangles, solid line)
at fixed value of $\gamma_2 = 0.3$ ($\sigma =0.3$).
The lines show the corresponding power laws.
% [use int.A02]
}
\end{figure}
The numerical results for the $x$-component of dimensionless force
$\delta \! f = \delta {\mathbf F}_x/(k_c u_0)$
are presented in Fig.~\ref{int.A02}.
The function $\delta \! f (r)$ exhibits a crossover from a slow Coulomb
like decay 
$\delta \! f (r) \propto r^{-1}$ at short distances $r \ll \lambda_c$
to the fast dipole-dipole like decay $\delta \! f (r) \propto r^{-3}$
at large distances $r \gg \lambda_c$.
The near and far zones are separates by the
\textit{elastic correlation length} $\lambda_c$
first introduced by Caroli and Nozieres~\cite{CN1998}.
It may be estimated in the following way:
the stiffness of the ``rigid block'' $K \sim E \lambda_c$
should be compensated by that of the interface,
$K \sim k_c \, (\lambda_c/a)^2$
(stiffness of one contact times the number of contacts).
This leads to
\begin{equation}
\label{int.eq30}
\lambda_c \approx a/ \gamma_1  = a^2 E/k_c \,.
\end{equation}

The rigid slider corresponds to the limit $E \to \infty$, or $\gamma_1 \to 0$.
Therefore, the slider may be considered as a rigid body
(e.g., in MD simulation),
if its size is smaller than $\lambda_c$.
For the steel slider considered in Sec.~\ref{material},
estimation gives $\lambda_c /a \sim 10^2$.
Up to distance $\lambda_c$ the contacts strongly interact.
If the $i$th contact breaks and its stretching changes on
$| \delta x_i | \approx x_c$,
then the force on the $j$th contact
at a distance $r_{ij} < \lambda_c$ % from the broken one,
away,
changes by
$\delta \! f_j \approx \tilde{\kappa} k_c a \, \delta \! x_i /r_{ij}$,
where the dimensionless parameter $\tilde{\kappa} < 1$
characterizes the strength of interaction
(numerics gives $\tilde{\kappa} \sim 10^{-3}$).
In the near zone $r \ll \lambda_c$
the interaction between the contacts may be accounted for
within the master equation approach
in a mean-field fashion as described in the next Sec.~\ref{MF}.
At larger distances, different regions of the slider
will undergo different displacements.
Therefore, in the far zone, $r \gg \lambda_c$,
we must take into account the elastic deformation of the slider.

% $K < K^*_{\rm eff}$ than \textit{screening} length.

% $K > K^*_{\rm eff}$ than \textit{crack}, solitonic wave.

%=========================================================================
\section{Nearby contacts: mean field approach}
\label{MF}

\begin{figure} % [h] %[t] \bigskip
\includegraphics[clip, width=8cm]{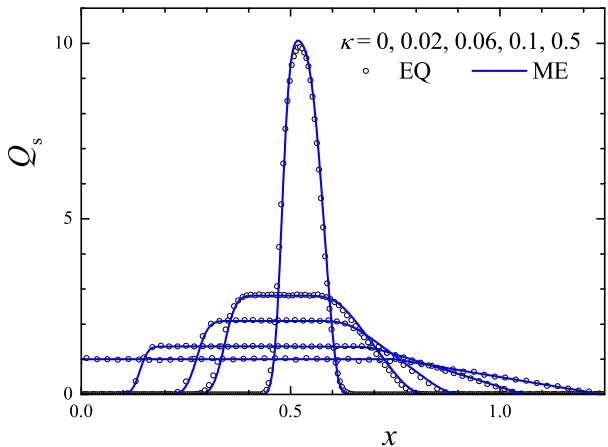}
\caption{\label{mf.B02a}(color online):
The steady state distribution $Q_s (x)$
for the rectangular threshold distribution $P_{c0} (x)$
with $x_s =1$ and $\Delta x_s =0.25$ and
different values of the interaction strength $\kappa =0$,
0.02, 0.06, 0.1, and 0.5.
The EQ simulations (dotted) are compared with
the ME results (solid curves).
% The fitting parameters are given in table~\ref{parameters} (variant~I).
%[use mf.B02a]
}
\end{figure}
\begin{figure} % [h] %[t] \bigskip
\includegraphics[clip, width=8cm]{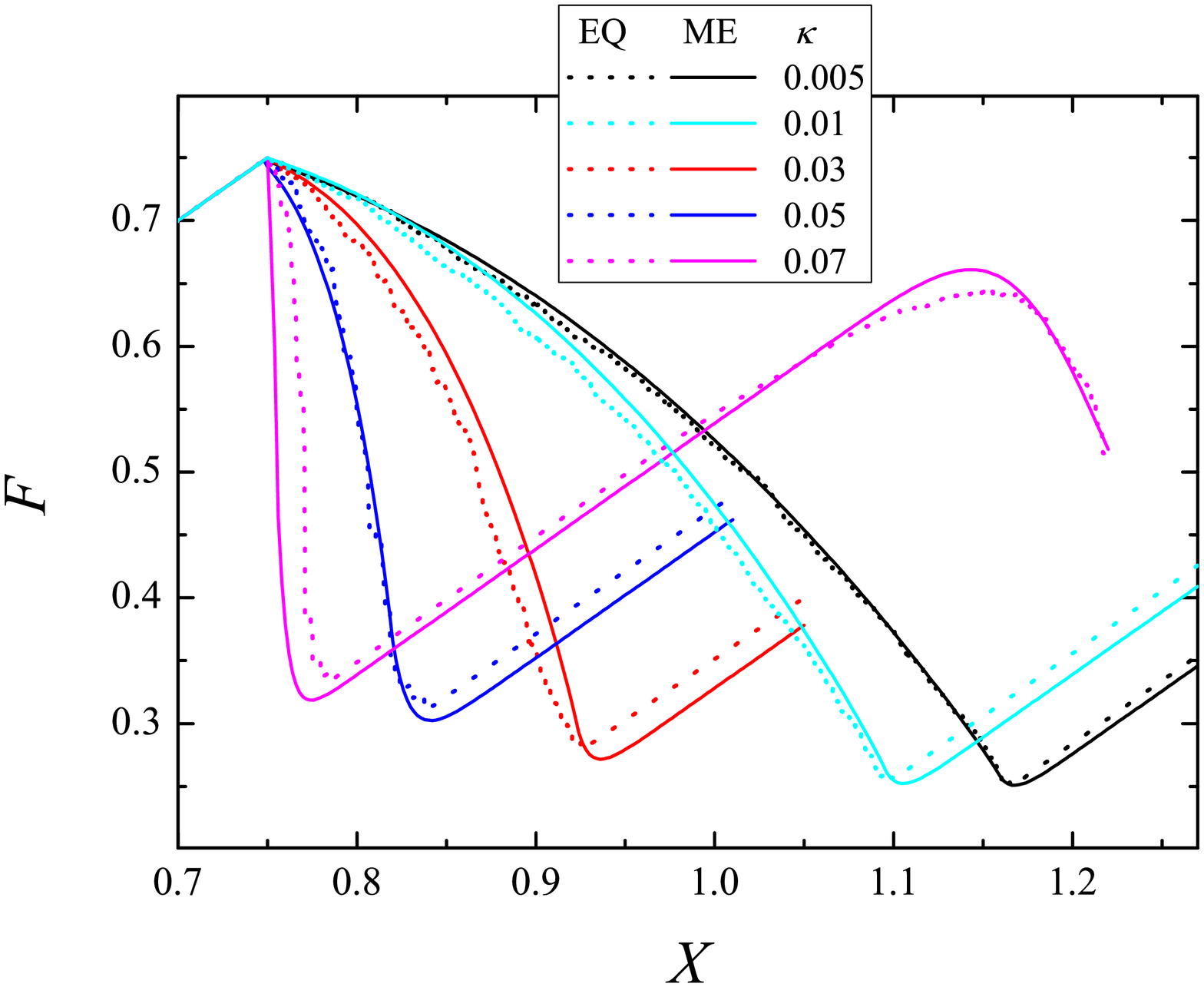}
\caption{\label{mf.B03}(color online):
Onset of sliding: the initial part of the dependence of
the friction force $F$ on the slider displacement $X$
for different strength of interaction
$\kappa =0.005$ (black),
0.01 (cyan), 0.03 (red),
0.05 (blue), and 0.07 (magenta).
Dotted curves show the results of EQ simulation, and
solid curves, the mean-field ME approach.
The threshold distribution $P_{c0} (x)$
has the rectangular shape with $x_s =1$ and
$\Delta x_s =0.25$. % (variant~I in table~\ref{parameters}).
%[use mf.B03]
}
\end{figure}
\textit{EQ model with interaction between the contacts}.
Let us now include the dynamical interaction between the contacts.
When a contact breaks,
the now unsustained shear stress
% sustained by this contact,
must be redistributed among the neighboring contacts.
% (and may even cause an avalanche of breaking contacts~\cite{BR2002}).
We assume that
because of elastic interaction between the contacts $i$ and $j$,
the forces acting on these contacts have to be corrected as
$f_i \to f_i - \Delta f_{ij}$ and $f_j \to f_j + \Delta f_{ij}$, where
$\Delta f_{ij} = k_{ij} \, (x_j - x_i)$
in linear approximation.
For example, let at the beginning the contacts be relaxed,
$x_j (0) = x_i (0) =0$. Due to sliding motion,
all stretchings grow together, so that still $\Delta f_{ij} = 0$.
At some instant $t$ let the $j$th contact break,
$x_j (t) \to 0$, with the $i$th contact still stretched, $x_i (t) >0$.
Clearly, as the $j$th contact breaks,
the force on the $i$th contact increases,
$\Delta f_{ij} (t) = - k_{ij} \, x_i (t) <0$.
The amplitude of interaction decreases with the distance $r$
from the broken contact as
$\Delta f \propto r^{-1}$
at short distances $r< \lambda_c$.
Neglecting the anisotropy of interaction, we assume that
$k_{ij} = \widetilde{f} / |r_{ij}|$,
where $\widetilde{f}$ is a parameter.

We simulated a triangular lattice of
$N=60 \times 68=4080$ contacts with periodic boundary conditions and
lattice constant $a=1$, with an average contact spring constant $k_c =1$
and radius of interaction $\lambda_c =3a$
% (when each contact interacts with 36 neighbors, variants~I, III and~IV)
or $\lambda_c =5a$.
% (88 neighbors, variant~II in table~\ref{parameters}).
We assumed $f_{bi}=0$ and a
rectangular shape of the distribution $P_c (x)$, i.e.,
$P_c (x) = P_{c0} (x) = (2 \Delta x_s)^{-1}$ for $|x-x_s|<\Delta x_s$
and 0 otherwise, % (see inset in Fig.~\ref{B01a}),
which admits an exact solution for noninteracting contacts~\cite{BP2010}
(more realistic distributions give the same results).

Figures~\ref{mf.B02a} and~\ref{mf.B03}
show the result of simulations for different values of the
dimensionless strength of the interaction
\begin{equation}
\kappa = \widetilde{f} /(k_c x_c) \,,
\label{mf.eqi14}
\end{equation}
where
% \begin{equation}
$x_c = \int dx \, x P_{c0} (x)$
% \label{mf.xc} \end{equation}
is the average stretching of the initial threshold distribution
(for the rectangular distribution $x_c =x_s$).
These results yield the following conclusions.
{\it First}, in the steady state, the interaction %results in shrinking
causes a narrowing of the final distribution $Q_s (x)$.
At high interaction strength $\kappa$, the distribution approaches 
a narrow Gaussian. % (see Fig.~\ref{B01a}).
{\it Second}, the drop of frictional force $F(X)$ at the onset of sliding
(at $X \sim x_c$) gets steeper and steeper as $\kappa$ grows.
Therefore, contact interactions reinforce elastic instability.
{\it Third}, above a critical interaction strength,
$\kappa \geq \kappa_c \sim 0.1$, %many
a multiplicity of contacts break simultaneously at the onset of sliding,
and there is an avalanche, where the force $F(X)$ drops abruptly.
% (see Fig.~\ref{B01b}b).
The average avalanche size %of such avalanches
may be estimated
similarly as done
in Ref.~\cite{BR2002}.

While the full EQ model may be only studied numerically, 
it is always useful to have % some
analytical results, even if only of qualitative level.
In what follows we show that the main EQ results
may be reproduced within the ME approach 
by using ``effective'' $P_c (x)$ and $R(x)$ distributions
defined in a mean-field fashion. In this section the ME equation is only
used to reproduce the EQ results. This is however useful because it
provides an additional understanding of the results as the 
effective distributions, obtained in this analysis, provide a description of
the collective effects affecting the contacts in terms of simple
functions. 

%-------------------------------------------------------------------------
\smallskip
\textit{Smooth sliding}.
Using the steady state solution of the ME,
Eqs.~(\ref{mf.eq07}) and~(\ref{mf.eq08}),
one may approximately recover
the functions $P_c (x)$ and $R(x)$
if the stationary distribution $Q_s (x)$ is known.
Indeed, for small $x$, where $P(x)$ is close to zero,
the left-hand side of $Q_s (x)$
allows us to find $R(x)$ as
$R(x) \propto Q^{\prime}_{s} (x)$ (see Eq.~(\ref{mf.eq06})),
while the right-hand side of $Q_s (x)$,
where $x \sim x_c$ and the contribution of $R(x)$
to the shape of the steady state distribution is negligible, 
gives us~\cite{BP2010}
$P_c (x) \propto P(x) \, Q_s(x) \propto -Q^{\prime}_{s} (x)$.
Thus, differentiating the function $Q_s (x)$
obtained in the EQ simulation,
we may guess shapes of the effective distributions $P_c (x)$ and $R(x)$
which, when substituted in the ME, would produce
a solution $Q_s (x)$ close to that obtained in the EQ simulation.

Using the simulation results, let us suppose that
the detached contacts form again with nonzero stretchings, i.e.,
that the distribution $R(x)$ is shifted to positive stretching values,
\begin{equation}
R(x) = G( x-\alpha x_c, \gamma x_c ) \,,
\label{mf.eq09}
\end{equation}
where $G(x, \sigma)$ is the Gaussian distribution
with zero mean and standard deviation~$\sigma$,
\begin{equation}
G(x, \sigma) =
\frac{1}{\sigma \sqrt{2\pi}}
\exp \left( - \frac{x^2}{2 \sigma^2} \right) .
\label{mf.eq10}
\end{equation}
At the same time,
we suppose that the effective threshold distribution $P_c (x)$
shrinks and shifts
with respect to the original (``noninteracting'') one,
\begin{equation}
P_h (x) = \beta P_{c0} \left[ \beta (x - \alpha x_c) \right] \,.
\label{mf.eq11}
\end{equation}
Let us moreover take its convolution with the Gaussian function~(\ref{mf.eq10}),
% \begin{equation}
$P_c (x) = P_h \otimes G
\equiv \int d\xi \, P_h (x - \xi) \, G(\xi, \gamma \sqrt{2} x_c)$.
% \label{mf.eq12} \end{equation}

The results of this procedure for the rectangular distribution $P_{c0} (x)$
are shown in Fig.~\ref{mf.B02a}.
We see that with a proper choice of
the parameters $\alpha$, $\beta$ and $\gamma$,
the ME solutions $Q_s (x)$ perfectly fits the numerical EQ results
(for the parameters $\alpha$, $\beta$ and $\gamma$
in Fig.~\ref{mf.B02a} we used expressions
% \begin{eqnarray} \label{eq17}
$\beta = 1 + b_1 \kappa$, % \,,
% \\ \label{eq18}
$\alpha = b_2 \kappa /\beta$ and % , % \,,
% \\ \label{eq19}
$\gamma = b_3 \alpha - b_4 \alpha^2$
% \end{eqnarray}
with the coefficients
$b_1 = 18$,
$b_2 = 9.6$,
$b_3 = 0.142$ and
$b_4 = 0.232$).
Results
of similar quality were also obtained for other simulated cases, e.g.,
for larger radius of the interaction or
for wider threshold distribution $P_{c0} (x)$.

The dependences of the fitting parameters $\alpha$, $\beta$ and $\gamma$
on the dimensionless strength of interaction $\kappa$
may be found in the following way. %Evidently,
To begin with, for noninteracting contacts initially
% it must be
$\alpha =\gamma =0$ and $\beta =1$.
It is reasonable to expect that in the lowest approximation
$\alpha, \gamma \propto \kappa$ and $\beta -1 \propto \kappa$.
Indeed, because the shift of the effective distribution$P_c (x)$
appears because of the interaction,
$\alpha f_c = \sum_j \Delta f_{ij}$\,, at small $\kappa$ we have approximately
\begin{equation}
\alpha \sim 0.5 \, a^{-2} \int_0^{\lambda_c} d^2 r \, \kappa x_c /|r|
= \pi \kappa {\lambda_c} x_c /a^2.
\end{equation}
At large $\kappa$, however, $\alpha$ has to saturate,
e.g., as $\alpha \propto \kappa /\beta$,
because the shift cannot be larger than $x_c$, i.e., $\alpha <1$.
Then, because the distribution $P_{c} (x)$
shrinks from both sides, we have $b_1 \sim 2 \, b_2$.

Thus, the interaction makes the threshold distribution
$P_c (x)$ narrower by a factor $\beta$
and shifts its center to the left-hand side,
$x_c \to \nu x_c$, where $\nu=\alpha + \beta^{-1}$
changes from 1 to 0.5 as the interaction strength $\kappa$ increases
from zero to infinity.

%-------------------------------------------------------------------------
\smallskip
\textit{Onset of sliding}.
The beginning of motion when started from the relaxed configuration,
$Q(x; \, 0) = \delta (x)$,
cannot be explained by the approach used above,
because the effective  distribution $P_c (x)$
is ``self-generated'' during smooth sliding, i.e.,
it can be applied only when
the process of contacts breaking--reattachment is continuously operating.
% es already.
%
Nevertheless, the initial part of the $F(X)$ dependence
may still be described by the effective ME approach,
but with the modified ``forward'' threshold distribution
given by the expression
\begin{equation}
P_{ci} (x) = {\cal N}
x^{\epsilon_0} P_{c0} \left[ \beta_0 (x - \alpha_0 x_c) \right] \,,
\label{mf.eq15}
\end{equation}
where ${\cal N}$ is a normalization factor,
$\int_0^{\infty} dx \, P_{ci} (x) =1$.
The parameter $\alpha_0$
is now defined so as to keep the lowest boundary unshifted,
$\beta_0 (x_{{\rm fix} \, 0} - \alpha_0 x_c) = x_{{\rm fix} \, 0}$
with $x_{{\rm fix} \, 0} = x_L = x_s - \Delta x_s$, so that
% \begin{equation}
$\alpha_0 = (x_{{\rm fix} \, 0} /x_c) (1 - \beta_0^{-1})$.
% \label{} \end{equation}
%
The ``backward'' distribution $R(x)$ is still defined by
Eq.~(\ref{mf.eq09}) with the same parameters as above.
% (the parameter $\gamma$ is not too relevant now;
% we used $\gamma = 2(b_3 \alpha - b_4 \alpha^2)$
% for the sake of concreteness).

Numerics shows that with a proper choice of the fitting parameters
$\beta_0$ and $\epsilon_0$ for a given value of $\kappa$,
the initial part of the function $F(X)$ may be reproduced
with quite high accuracy.
Moreover, for a rather wide range of $\kappa$ values,
the EQ simulation results may be reproduced by the ME approach
with a reasonable accuracy
using only three fitting parameter $c_1$, $c_2$ and $\kappa_c$,
if the parameters $\beta_0$ and $\epsilon_0$ in Eq.~(\ref{mf.eq15})
are given by the expressions
% \begin{eqnarray}
$\beta_0 = 1 + c_1 \kappa /(1 - \kappa /\kappa_c )$ and % \,,
% \label{eq15a} \\
$\epsilon_0 = c_2 (\beta_0 -1)$, %  \,,
% \label{eq15b} \end{eqnarray}
where the parameter $\kappa_c$ corresponds
to the critical ``breakdown'' interaction strength when many contacts
begin to break simultaneously.
For $\kappa > \kappa_c$,
the drop of $F(X)$
% after the first maximum changes to a
becomes jump-like, so that
$K^* = \infty$ and stick-slip will appear for any
stiffness of the slider $K < \infty$.
Note that the value of $\kappa_c$ may be estimated from the equation
$\alpha x_c \sim \Delta x_s$.

For the rectangular shape of the distribution $P_{c0} (x)$
% with the parameters $x_s =1$ and $\Delta x_s = 0.25$ used above,
the result of this procedure is demonstrated in Fig.~\ref{mf.B03}
(the fitting parameters are
$c_1 = 33.9$, $c_2 = 3.0$ and $\kappa_c = 0.074$).

Of course, the $P_{ci} (x)$ function, Eq.~(\ref{mf.eq15}), can describe only
the initial part of the $F(X)$ dependence,
when $F(X)$ grows, reaches the first maximum and then decreases.
To simulate the whole dependence $F(X)$,
one would have to involve the evolution of $P_c (x)$ with sliding distance,
e.g., as some ``aging'' process $P_{ci} (x) \to P_{c} (x)$
(see Ref.~\cite{BP2010})
with the initial distribution
$P_{c, \, \rm ini} (x) = P_{ci} (x)$
and the final one
$P_{c, \, \rm fin} (x) = P_c (x)$.

\begin{figure} % [h] %[t] \bigskip
\includegraphics[clip, width=8cm]{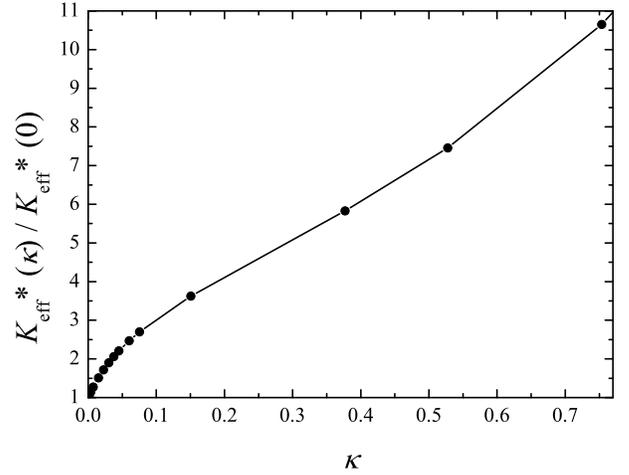}
\caption{\label{mf.B08} % (color online):
The effective interface stiffness
$K_{\rm eff}^*$
(normalized on the noninteracting value)
as a function of the strength of interaction $\kappa$
for the realistic threshold distribution
$P_{c0} (x) = (2/ x_s) \, u^3 e^{-u^2}$, $u \equiv x/x_s$ with $x_s =1$,
when $K^* /K_s = 0.179$.
%[use mf.B08]
}
\end{figure}
\smallskip
\textit{Stick-slip versus smooth sliding}.
As was mentioned above, stick-slip appears as a result of elastic instability
which is controlled by the relation
between the slider stiffness $K$
and the effective interface stiffness $K^*$.
For noninteracting contacts
$K^* \approx K_s x_c / \Delta x_s$;
because typically $\Delta x_s \sim x_c$,
estimates give $K^* \alt K$
so that stick-slip should never appear.
% as long as the interaction between the contacts is ignored.
The interaction between contacts strongly enhances the elastic instability
thus making stick-slip much more probable.
Indeed, because of the effective shrinking of the threshold
distribution, the parameter 
$K^*$ increases roughly as $K^* \to K_{\rm eff}^* \sim \beta_0 K^*$, i.e.,
the effective interface stiffness
$K_{\rm eff}^*$ grows with the strength of interaction $\kappa$,
and the elastic instability can now appear.
For example, for a realistic threshold distribution
the dependence of $K_{\rm eff}^*$
on the strength of interaction $\kappa$
is shown in Fig.~\ref{mf.B08}.

The strength of interaction between the contacts may be found as
$\kappa \approx \bar{\kappa} a/x_c$, where
realistic values of the dimensionless parameter $\bar{\kappa}$
are of the order $\bar{\kappa} \sim 10^{-3}$;
taking $a \sim (10^2 \div 10^3) \, r_c$ and $x_c \sim r_c$,
we obtain $\kappa \sim 0.1 \div 1$ %.
% Then, using $b_1 \approx 2\pi \lambda_c x_c /a^2$, we obtain that
% $\beta_0 \sim \beta \sim 2\pi \bar{\kappa} \lambda_c /a$.
which gives $\beta_0 \sim 3 \div 13$ according to Fig.~\ref{mf.B08}.

%=========================================================================
\section{Far zone: Meso/macroscale friction}
\label{farzone}

At the {\it mesoscopic scale}, i.e.\ on distances $r \gg \lambda_c$,
the substrate must be considered as deformable.
Let us split the frictional area into (rigid) blocks of size $\lambda_c$.
In a general 3D model of the elastic slider,
the $n$th $\lambda_c$-block is
characterized by a coordinate $X_n$, and its dynamics
% (we explore below the scalar variant of the model),
is described by the ME
for the distribution functions $Q_n (u_n; X_n)$.
A solution of these MEs gives the interface forces $F_n (X_n)$.
% as described in Ref.~\cite{BP2010}.
Then, the transition from the discrete numbering of blocks
to a continuum interface coordinate $r$ is trivial:
$n \to r$,
$Q_n (u_n; X_n) \to Q[u; X(r); r]$,
$P_n (u) \to P(u; r)$,
$\Gamma_n (X_n) \to \Gamma [X(r); r]$,
$F_n (X_n) \to F[X(r); r]$
(here $r$ is a two-dimensional vector at the interface),
and the master equation now takes the form:
\begin{widetext}
\begin{equation}
\frac{\partial Q[u; X(r); r]}{\partial X(r)} +
\frac{\partial Q[u; X(r); r]}{\partial u} +
P(u) \, Q[u; X(r); r] =
\delta (u) \, \Gamma [X(r); r] \,,
\label{crack.eq03x}
\end{equation}
\end{widetext}
where we assumed that, for the sake of simplicity,
the contacts are reborn with zero stretching, $R(u) = \delta(u)$,
and
\begin{equation}
\Gamma [X(r); r] = \int d\xi \, P( \xi ) \, Q[\xi; X(r); r] \,.
\label{crack.eq05x}
\end{equation}
Equations (\ref{crack.eq03x},~\ref{crack.eq05x})
should be completed with the elastic equation of motion
for the sliding body (we assume isotropic slider)
\begin{equation}
\ddot{\mathbf u} + \eta \dot{\mathbf u} =
G_1 \nabla^2 {\mathbf u} + G_2 \nabla (\nabla \cdot {\mathbf u} ) \,,
\label{crack.eq06}
\end{equation}
where ${\mathbf u} ({\mathbf R})$ is the 3D displacement vector
in the slider (${\mathbf R} = \{ x,y,z \}$),
$\eta$ is the intrinsic damping in the slider,
$G_1 = E/2(1+\sigma) \rho = c_{t}^2$ and
$G_2 = G_1 /(1-2\sigma) \rho = c_l^2 -c_t^2$,
$E$, $\sigma$ and $\rho$ are the Young modulus,
Poisson ratio and mass density of the slider correspondingly,
and $c_l$ ($c_t$) is the longitudinal (transverse) sound speed.
Equation~(\ref{crack.eq06}) should be solved
with corresponding boundary and initial conditions.
In particular, at the interface (the bottom plane of the slider,
where $z=0$ and $\{ x,y \} = r$)
we must have ${\mathbf u}_x = X(r)$,
${\mathbf u}_y = {\mathbf u}_z =0$, and
the shear stress should equal  $F[X(r); r]/\lambda_c^2$,
where the friction force acting on the $\lambda_c$-block from the interface,
\begin{equation}
F[X(r); r] = N_{\lambda} k_c \int du \, u \, Q[u; X(r); r] \,,
\label{crack.eq07}
\end{equation}
should be obtained from the solution of Eq.~(\ref{crack.eq03x})
[here $N_{\lambda} = (\lambda_c /a)^2$].

Equations~(\ref{crack.eq03x}--\ref{crack.eq07}) form the complete set of
equations 
which describes evolution of the large scale tribological system;
in a general case it has to be solved numerically.
However, a qualitative picture may be obtained analytically.
The interface dynamics depends on whether or not the $\lambda_c$-blocks
undergo the elastic instability, i.e., on the ratio
of the stiffness of the $\lambda_c$-block
$K_{\lambda} \approx (2 c_l^2 + 3 c_t^2) \rho \lambda_c$
[as follows from the discretized version of
Eq.~(\ref{crack.eq06})]
and the effective critical stiffness parameter of the interface
$K^*_{\lambda \; \rm eff} = \beta_0 K^*_{\lambda}$,
where
$K^*_{\lambda} \sim K_{\lambda s} x_c /\Delta x_s$ and
$K_{\lambda s} = N_{\lambda} k_c$.
If the elastic instability does not appear, then a local perturbation
at the interface relaxes, spreading over an area of size $\lambda_s$ ---
the screening length considered below in Sec.~\ref{screening}.
In the opposite case, when the elastic instability does emerge (locally),
in may propagate through the interface.
Below in Sec.~\ref{crack} we consider a simplified one-dimensional version,
which allows us to get some analytical results and a rather simple
simulation approach 
(such a model is also supported by the fact that
the largest forces near the broken contact
are just ahead/behind it according to Fig.~\ref{int.A10}).
Recall that the interaction between the $\lambda_c$-blocks is weaker than
in the short-range zone, it follows the law $\delta \! f \propto r^{-3}$
which determines, e.g., the block-block interaction strength $\kappa_{\lambda}$
in Eq.~(\ref{scr.eq01Pi}) below
(although the interaction is power-law,
% i.e., formally it is long-ranged,
we may consider
nearest neighbors only, because excitations at the interface,
such as ``kinks'' introduced in Sec.~\ref{crack},
are localized excitations, and the role of long-range character
of the interaction reduces to modification of their parameters~\cite{BKbook}).

%-------------------------------------------------------------------------
\subsection{Elastic screening length}
\label{screening}

Let us assume that the slider is split in $\lambda_c$-blocks (rigid blocks)
and consider the block-block interaction in a mean-field fashion
(analogously to methods used in soft matter,
see Refs.~\cite{S1998,HL1998,BCA2009}).
Due to sliding of neighboring blocks,
the forces acting on contacts in
the $n$th $\lambda_c$-block get an addition shift.
This effect may be accounted with the help of a substitution
$f_n \to f_n + \Delta f_n$,
$\Delta f_n = \sum_{m \neq n}
f_m \times {\rm Prob(}m \to {\rm broken)} \times \Pi_{mn}
\approx x_c \sum_{m \neq n} f_m \, \Gamma_m \Pi_{mn}$
(recall that the sum is over the $\lambda_c$-blocks here),
or approximately
\begin{equation}
f_n \to \left[ 1 + x_c \sum_{m \neq n} \Gamma_m (X_m) \, \Pi_{mn} \right] f_n \,,
\label{scr.eq01}
\end{equation}
where
$\Gamma_m (X_m) = \int du \, P_m (u) \, Q_m (u; X_m)$ so that
$N_{\lambda} \Gamma_m (X_m)$
is the number of broken contacts in the $m$th $\lambda_c$-block
per its unit displacement,
and
\begin{equation}
\Pi_{mn} \approx N_{\lambda} \kappa_{\lambda} \, (\lambda_c /r_{mn})^{3}
\label{scr.eq01Pi}
\end{equation}
describes the dimensionless (i.e., normalized on $f_s$)
elastic interaction between the $\lambda_c$-blocks
separated by the distance $r_{mn}$.
In this way the force is given by $f_s \Pi$; the numerical constant
$\kappa_{\lambda} \sim \bar{\kappa} a/ \lambda_c$
depends on the substrate and interface parameters.

Let us introduce the dimensionless variable
$\varepsilon_n = x_c \sum_{m \neq n} \Gamma_m (X_m) \, \Pi_{mn}$.
The shift of forces in the $n$th block
due to broken contacts in the neighboring blocks
may be accounted by a renormalization of the rate:
\begin{equation}
P_n (u) \to P_n \left[ (1+\varepsilon_n) u \right] \,.
\label{scr.eq02}
\end{equation}
Indeed, when contacts in the neighboring blocks break,
then the forces in the given block increase, $\varepsilon_n > 0$,
and the contacts in the given block should start to break earlier, i.e.,
their threshold distribution effectively shifts to lower values.

Making the transition from discrete sliding blocks
to a continuum sliding interface,
$\Pi_{mn} \to \Pi (r'-r)$ and
$\varepsilon_n \to \varepsilon (r)$,
we obtain a master equation of the form:
\begin{widetext}
\begin{equation}
\frac{\partial Q[u; X(r); r]}{\partial X(r)} +
\frac{\partial Q[u; X(r); r]}{\partial u} +
P\left( [1+\varepsilon (r) ] u \right) \, Q[u; X(r); r] =
\delta (u) \, \Gamma [X(r); r] \,,
\label{scr.eq03}
\end{equation}
\end{widetext}
where we again assumed that %, for the sake of simplicity,
the contacts are reborn with zero stretchings, $R(u) = \delta(u)$,
\begin{equation}
\varepsilon (r) = x_c \lambda_c^{-2}
\int_{|r' -r| \geq \lambda_c} d^2r' \, \Gamma [X(r'); r'] \, \Pi (r'-r) \,
\label{scr.eq04}
\end{equation}
and
\begin{equation}
\Gamma [X(r); r] = \int d\xi \, P\left( [1
+\varepsilon (r)] \xi \right) \, Q[\xi; X(r); r] \,.
\label{scr.eq05}
\end{equation}

In the long-wave limit, when
$|d\varepsilon (r)/dr| \ll \varepsilon (r)/\lambda_c$,
we may assume that the interface is locally equilibrated,
i.e., the distribution of forces on contacts is close to the
steady-state solution 
of the master equation, $Q[u; X(r); r] \approx Q_s (u; r)$,
which depends parametrically on the coordinate $r$ through the function
$\varepsilon (r)$ entered into the expression for the rate
$P\left( [(1+\varepsilon (r)] u \right)$.
The stationary solution of the ME is known analytically~\cite{BP2010},
and we may find the function~(\ref{scr.eq05}),
$\Gamma (r) = [1 +\varepsilon (r)]/x_c$.
Together with Eq.~(\ref{scr.eq04}) this gives a self-consistent equation
on the function $\varepsilon (r)$:
\begin{equation}
\varepsilon (r) = \lambda_c^{-2}
\int_{|r' -r| \geq \lambda_c} d^2r' \, [(1+\varepsilon (r')] \, \Pi (r'-r) \,.
\label{scr.eq08}
\end{equation}

Taking into account the interaction
of nearest neighboring $\lambda_c$-blocks only
and expanding $\varepsilon (r)$ in Taylor series,
we obtain the equation
\begin{equation}
\varepsilon (r) = \Pi_0 \left[1+ \varepsilon (r) +
{1\over 2} \, \lambda_c^2 \varepsilon'' (r) \right] ,
\label{scr.eq09}
\end{equation}
where
$\Pi_0 = \nu \Pi (\lambda_c) = \nu N_{\lambda} \kappa_{\lambda}
\sim \nu \bar{\kappa} \lambda_c /a$ and
$\nu = 2 \div 4$ is the number of nearest neighbors.
Writing $\varepsilon (r) = \varepsilon_0 + \Delta \varepsilon (r)$,
where $\varepsilon_0 = \Pi_0 / (1- \Pi_0)$,
Eq.~(\ref{scr.eq09}) may be rewritten as
\begin{equation}
\lambda_s^2 \Delta \varepsilon'' (r) = \Delta \varepsilon (r) \,,
\label{eq10}
\end{equation}
where $\lambda_s = \lambda_c (\varepsilon_0 /2)^{1/2}$
is the characteristic screening length in the sliding interface.

From the known analytical steady state solution of the ME ~\cite{BP2011},
we may predict the dependence of screening length
on temperature and sliding velocity. In particular,
if $T>0$, then
$\lambda_s \propto v^{-1/2} \to \infty$ as $v \to 0$
in agreement with the results of Ref.~\cite{LC2009}.

%-------------------------------------------------------------------------
\subsection{Frictional crack as a solitary wave}
\label{crack}

In the frictional interface, sliding begins at some weak place
and then expands throughout the interface.
Such a situation is close to the one known in fracture mechanics
as the mode II crack, when the shear is applied along the fracture plane.
In friction, a crack first opens,
evolves (propagates, grows, extends) during some ``delay'' time~$\tau$,
but then it either %should
expands throughout the whole interface,
or it will {\it close\/} because of the load. % (Fig.~\ref{A01}b).
Below we consider the latter scenario,
when one solid slips over another
due to motion of the so-called self-healing
crack~\cite{C2000,GM2001,G2004,G2007} 
--- a wave or ``bubble'' of separation moving like a crease on
rug~\cite{VBA2009}. 
Our plan is to adopt ideas from fracture mechanics,
adapt them to the friction problem and then reduce it
to the Frenkel-Kontorova (FK) model~\cite{BKbook}
in order to describe collective motion of contacts in the frictional interface.

When one of the ``collective contacts''
(the $\lambda_c$-block) breaks, it may initiate
a chain reaction, with contacts breaking domino-like one after another.
This scenario may be described accurately by reducing
the system of contacts to a Frenkel-Kontorova-like model. % ~\cite{BKbook}.
Recall that the FK model describes
a chain of harmonically interacting atoms subjected to
the external periodic potential  $V_{\rm sub} (x)$ of the substrate.
% (see Fig.~\ref{A01}d).
If the atoms are additionally driven by an external force~$f$,
then the equations of motion for the atomic coordinates $u_n$ take the form
\[
m \ddot{u}_n + m \eta \dot{u}_n - g (u_{n+1} + u_{n-1} -2 u_n)
+ V^{\prime}_{\rm sub} (u_i) = f \,,
\]
where $m$ is the atomic mass, $g$ is the strength of elastic interaction
between the atoms, and $\eta$ is an effective damping coefficient
which describes dissipation phenomena
such as the excitation of phonons {\it etc}.\ in the substrate.
The main advantage of using the FK model is that
its dynamics is well documented~\cite{BKbook}.
Mass transport along the chain is carried
by kinks (antikinks) ---
local compressions (extensions) of otherwise commensurate structure.
The kink is a well-defined topologically stable excitation
(quasiparticle) characterized by an effective mass
$m_k$ which depends on the kink velocity $v_k$,
$m_k = m_{k0} ( 1-v_k^2 /c^2 )^{-1/2}$
(the relativistic Lorentz contraction of the kink width
when its velocity approaches the sound speed $c$).
Therefore, the maximal kink velocity $v_{k \, \rm max} =c$.
In the discrete chain, kinks move in the so-called Peierls-Nabarro
(PN) potential, whose amplitude %of which
is much lower
than that of the primary potential $V_{\rm sub} (x)$.
Therefore, the kink motion is activated over these barriers, and
its minimal velocity $v_{k \, \rm min}$ is nonzero.
The steady-state kink
motion is determined by the energy balance:
the incoming energy (because of action of the external driving force~$f$)
should go to creation of new ``surfaces'' (determined by the amplitude
of the substrate potential)
plus excitation of phonons by the moving kink
(described by the phenomenological damping coefficient~$\eta$),
so that $v_k (f)=f/(m_k \eta)$.

\smallskip
%-------------------------------------------------------------------------
\textit{FK-ME model}.
Thus, let us consider a chain of $\lambda_c$-contacts
(``atoms'' of mass $m = \rho \lambda_c^3$),
coupled harmonically with an elastic constant $g$,
driven externally through a spring of elastic constant $K$
with the end moving with a velocity $v$.
Using the discretized version of
Eq.~(\ref{crack.eq06}), the elastic constants may be estimated as
$g \approx 2 \lambda_c \rho c_l^2$ and
$K \approx \lambda_c \rho c_t^2$.
The $\lambda_c$-contacts are
coupled ``frictionally'' with the bottom substrate;
the latter is described by the nonlinear force $F_s (u)$.
% (see Fig.~\ref{A01}e).
The equation of motion of the discrete chain is
\begin{equation}
m \ddot{u}_n + m \eta \dot{u}_n - g (u_{n+1} + u_{n-1} -2 u_n)
+ F_s (u_n) + K u_n = f \,,
\label{crack.eqi00}
\end{equation}
where the driving force is given by $f(t) = K v t$.
\begin{figure} % [h] %[t] \bigskip
\includegraphics[clip, width=8cm]{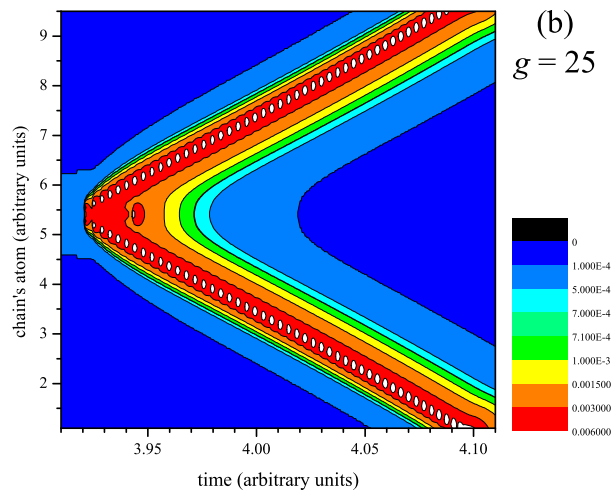}
\caption{\label{crack.A03ab}(color online):
Color map of atomic velocities for
a typical evolution of the chain of contacts.
The nearest neighboring contacts interact elastically with the constant $g=25$.
The interaction with the substrate is modeled by the function
$F_s (u) = k_c [\tanh (u) + 1.5 e^{-u} \sin (3u)]$ with $k_c =1$
defined for $0 \leq u < u_c =1$
and periodically prolonged for other values of $u$.
All contacts are driven through the springs of the elastic constant $K=0.07$,
their ends moving with the velocity $v =10^{-4}$.
The motion is overdamped ($m=1$, $\eta =100$).
To initiate the breaking, two central contact interact with the substrate
with smaller values of the elastic constant, $k_c =0.5$.
%[use crack.A03b]
}
\end{figure}
The substrate force $F_s (u)$ is found from the solution
of the ME for % the EQ-like model of
the rigid $\lambda_c$-block.
A typical evolution of the chain is shown in Fig.~\ref{crack.A03ab}.

The general case may only be investigated numerically.
Let us first consider a simplified case, when $F_s (u)$
has the sawtooth shape, i.e.\ it is defined as
\begin{equation}
F_s (u) = k_c u \;\;\; {\rm for} \;\;\; 0 \leq u < u_c
\label{crack.eqi08}
\end{equation}
and periodically prolonged for other values of $u$.
We assume that $f$ is approximately constant during kink motion
(otherwise, the kink will accelerate during its motion along the chain);
this is correct if the change of the driving force
$\Delta f = K v \, \Delta t$ during kink motion through the chain,
$\Delta t = L/v_k$ ($L$ is the chain length and $v_k$ is kink velocity),
is much lower than $k_c u_c$, or
% \begin{equation}
$K/k_c \ll (v_k/v)(u_c/L)$.
% \label{...} \end{equation}

Let us define the function
% \begin{equation}
${\cal F} (u) = F_s (u) + K u - f$.
% \label{...} \end{equation}
The degenerate ground states of the chain are  determined by the equation
${\cal F} (u) = 0$.
Let the right-hand side ($n \to \infty$) of the chain be unrelaxed,
$k_c u_R + K u_R = f$, or
\begin{equation}
u_R = f/(k_c +K) \,,
\label{crack.eqi09}
\end{equation}
while the left-hand side ($n \to -\infty$) already undergone relaxation,
$k_c (u_L - u_c) + K u_L = f$, or
\begin{equation}
u_L = (f+k_c u_c)/(k_c +K) \,.
\label{crack.eqi10}
\end{equation}
% as shown in Fig.~\ref{A01}e.

Thus, the FK-like model of friction (the FK-ME model) is described
by Eqs.~(\ref{crack.eqi00}) and~(\ref{crack.eqi08})
with the boundary conditions given by
Eqs.~(\ref{crack.eqi09}) and~(\ref{crack.eqi10}).

\smallskip
%-------------------------------------------------------------------------
\textit{Continuum-limit approximation}.
Let the system be overdamped ($\ddot{u}=0$);
later on we shall remove this restriction.
In the continuum-limit approximation,
$n \to x=na$ ($a=1$),
% but we shall keep it in formulas to control dimensions),
the motion equation takes the form
\begin{equation}
m \eta u_t - a^2 g u_{xx} + {\cal F} (u) =0,
\;\;\;
{\cal F} (u)|_{x \to \pm \infty} = 0 \,.
\label{crack.eqi01}
\end{equation}

We look for a solution in the form of
a wave of stationary profile (the solitary wave),
$u (x,t) = u(x-v_k t)$, so that
$u_t = -v_k u'$ and $u_{xx} = u''$.
In this case Eq.~(\ref{crack.eqi01}) takes the form
\begin{equation}
m \eta v_k u' + a^2 g u'' = {\cal F} (u) \,,
\label{crack.eqi04h}
\end{equation}
which may be solved analytically by standard methods~\cite{BP2011b}.

A solution of Eq.~(\ref{crack.eqi04h})
with these boundary conditions exists only
for a certain value of the kink velocity $v_k$,
defined by the equation
\begin{equation}
(m \eta v_k)^2 = ga^2 (k_c +K) {(2- \beta)^2}/{(\beta -1)} \,,
\label{crack.eqi04}
\end{equation}
where $\beta = k_c/(k_* -K)$ and $k_* = f/u_c$.
The solitary-wave solution exists for forces
$f_{\rm min} < f < f_{\rm max}$ only.
The minimal force which supports the kink motion ---
the Griffith threshold --- is given by
\begin{equation}
f_{\rm min} = \left( {1 \over 2} k_c +K \right) u_c \,.
\label{...}
\end{equation}
The maximal force, for which a kink may exist, is given by
\begin{equation}
f_{\rm max} = (k_c +K) \, u_c \,;
\label{...}
\end{equation}
at higher forces, the barriers of $F_s (u)$ are degraded,
the stationary ground states disappear,
and the whole chain must switch to %has to be in
the sliding state.

From Eq.~(\ref{crack.eqi04}) we can find the kink velocity
as a function of the driving force.
At low velocities
\begin{equation}
v_k \approx (f-f_{\rm min})/m_k \eta \,,
\label{crack.v1}
\end{equation}
where we introduced the effective kink (crack) mass
\begin{equation}
m_k = m \biggr/ \frac{4a}{u_c}
\sqrt{\frac{g}{k_c} \left( 1+\frac{K}{k_c} \right)} \,,
\label{crack.v1a}
\end{equation}
while at $f \to f_{\rm max}$ the velocity tends to infinity,
\begin{equation}
m \eta v_k \approx \sqrt{\frac{gk_c (k_c +K)a^2 u_c}{(f_{\rm max} -f)}} \;.
\label{...}
\end{equation}
The latter limit should be corrected by taking into account inertia effects.
The term $m \ddot{u}$ in Eq.~(\ref{crack.eqi00}) gives $mv_k^2 u''$
for the solitary-wave solution,
so it can be incorporated if we substitute in the above equations
$g \to g_{\rm eff} = g (1-v_k^2/c_0^2)$, where
$c_0 = (ga^2 /m)^{1/2}$ is the sound speed along the chain.
% (the % well-known ``relativistic'' narrowing of kinks).
The high-velocity limit now takes the form
\begin{equation}
v_k \approx c_0 \biggr/\sqrt{1+\frac{m \eta^2 (f_{\rm max} -f)}{k_c (k_c
    +K)u_c}} \;. 
\label{...}
\end{equation}

\begin{figure} % [h] %[t] \bigskip
\includegraphics[clip, width=8cm]{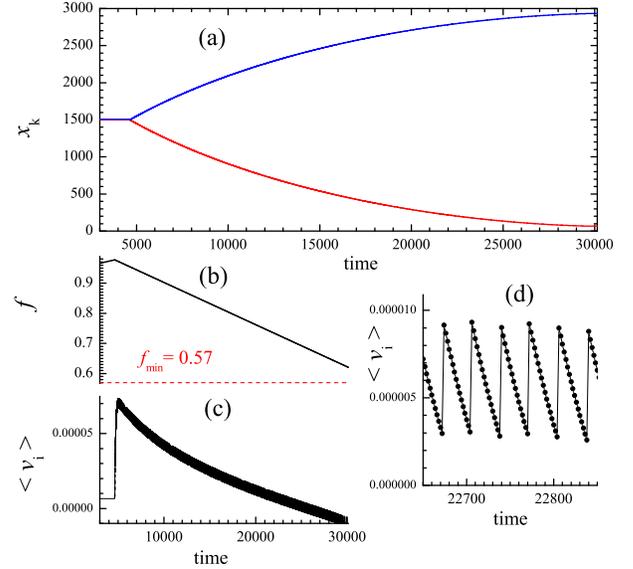}
\caption{\label{crack.A04a}(color online):
Evolution of the chain of $N=3000$ contacts.
The nearest neighboring contacts interact elastically with the constant $g=25$,
the interaction with the substrate is modeled
by the sawtooth function~(\ref{crack.eqi08}) with $k_c =1$ and $u_c =1$.
All contacts are driven through the springs of the elastic constant $K=0.07$,
their ends moving with the velocity $v =10^{-4}$.
The motion is overdamped ($m=1$, $\eta =100$).
To initiate the breaking, two central contacts interact with the substrate
with smaller spring constants, $k'_c=0.5$.
% ; also, to accelerate simulation, we used $x_{\rm ini}=13.5$.
When the kinks motion begins, % at $t=t_b$,
the elastic constants of the central contacts
restore their values to $k_c =1$,
and the driving velocity changes its sign,
$v \to v_b = -2 \times 10^{-4}$.
(a)~shows the kinks centers
(defined as places where the atomic velocity is maximal),
(b)~shows the driving force $f(t)$, % Eq.~(\ref{crack.eqi00a}),
(c)~shows the average chain velocity
$\langle \dot{u}_i \rangle = N^{-1} \sum_i \dot{u}_i$, and
(d)~demonstrates oscillation of the velocity due to PN barriers.
%[use crack.A04a]
}
\end{figure}
\smallskip
%-------------------------------------------------------------------------
\textit{Simulations}.
\begin{figure} % [h] %[t] \bigskip
\includegraphics[clip, width=8cm]{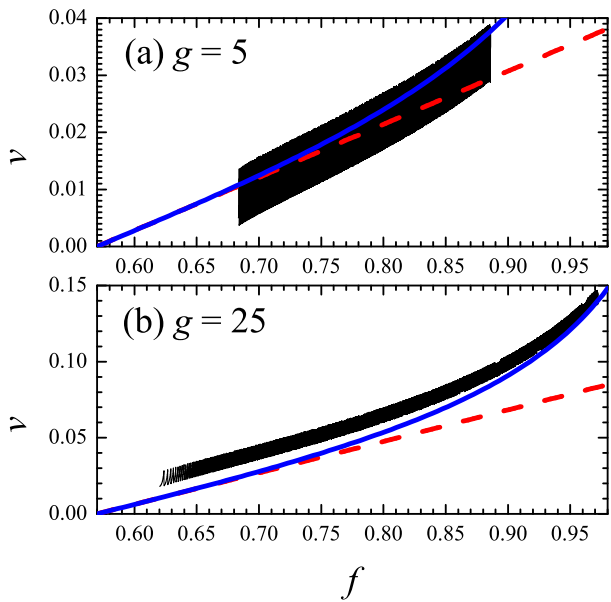}
\caption{\label{crack.A05}(color online):
Kink velocity versus the driving force for
(a)~$g=5$ ($v_b = -4 \times 10^{-5}$) and
(b)~$g=25$ ($v_b = -2 \times 10^{-4}$);
$N=3 \times 10^4$,
other parameters as in Fig.~\ref{crack.A04a}.
Blue solid and red dashed lines correspond to Eqs.~(\ref{crack.eqi04})
and~(\ref{crack.v1}), correspondingly.
% [use crack.A05]
}
\end{figure}
The continuum-limit approximate is accurate
for the case of strong interaction
between the contacts, $g \gg 1$;
in the opposite limit one has to %use
resort to computer simulation.
We solved Eq.~(\ref{crack.eqi00}) by the Runge-Kutta method.
As the initial state, we took the chain of length $N$
(typically $N=3 \times 10^3$ or $3 \times 10^4$)
with periodic boundary conditions and all contacts relaxed,
but the threshold breaking value for two central contacts was
taken lower than for the other contacts.
Then the driving force increases because of stage motion,
two central contacts break first and initiate two solitary waves
of subsequent contact breaking which propagate in the opposite directions
through the chain.
The value $k'_c$ of the lower threshold of the central contacts
determines the driving force and therefore the kink velocity;
the lower this threshold, the lower the threshold force 
for the motion to start \cite{BP2011b}.
As soon as the kink motion is initiated,
the $k_c$-values of the central contacts are restored to the same value
as for other contacts
(otherwise these contacts will act as a source of creation of new pairs
of kinks), 
and we begin to move the stage in the opposite
direction, $v >0 \to v_b <0$, so that
the driving force linearly decreases with time (see Fig.~\ref{crack.A04a}b),
the average chain velocity
$\langle \dot{u}_i \rangle = N^{-1} \sum_i \dot{u}_i$
decreases as well (Fig.~\ref{crack.A04a}c)
until the motion stops (Fig.~\ref{crack.A04a}a).
Also, such an algorithm allows us to find the dependence
of the kink velocity determined as
\begin{equation}
v_k = n_k^{-1} \, N \left( \langle \dot{u}_i \rangle - \bar{v} \right),
\label{...}
\end{equation}
where $n_k =2$ is the number of moving kinks in the chain
and $\bar{v} = \dot{u}_{L, R} = v_b K/(k_c +K)$ is the background velocity,
on the driving force $f$.
These dependences are presented in Fig.~\ref{crack.A05};
they agree well with that predicted
by Eqs.~(\ref{crack.eqi04}) and~(\ref{crack.v1}).

Contrary to the continuum-limit approximation,
in the discrete chain of contacts the kink oscillates during motion
(see Fig.~\ref{crack.A04a}d)
--- the well-known discreteness effect of the FK model
due to existence of the PN barriers $f_{\rm PN}$.
The stronger the elastic interaction between the contacts,
the larger the kink ``width''
and the smaller the kink oscillations
(compare Figs.~\ref{crack.A05}a and~~\ref{crack.A05}b).
The amplitude of oscillations also depends on the shape
of the ``substrate potential''~\cite{BKbook} ---
it is larger for a sawtooth potential $F_s (u)$,
but smaller for a smoother shapes.
Recall that the $\lambda_c$-contacts are characterized by
a smooth dependence $F_s (u)$ as follows from the master equation.
The PN oscillations determine the lowest average kink velocity.
Therefore, the lowest velocity allowed for the frictional crack propagation,
$v_{k \; \rm min}$,
is determined by the parameters $g$ and $\lambda_c$ ---
the larger are $g$ and $\lambda_c$, the smaller is $v_{k \; \rm min}$.

\smallskip
%-------------------------------------------------------------------------
\textit{Discussion}.
The FK-ME model used here is rather close to the well-known
1D Burridge--Knopoff (BK) model of earthquakes
with a velocity--weakening friction law~\cite{BK1967}.
The difference is in the interface force $F_s (u)$:
we use the function derived from the ME-EQ model
(with well-defined parameters which may be extracted from experiments
or calculated from first principles),
% and additionally may incorporate aging effects, \textit{etc.}),
whereas the BK % Burridge--Knopoff
model adopts a phenomenological
velocity-dependent function for $F_s$.
Nevertheless, the qualitative behavior of the two models is similar,
the BK model also exhibits solitary-wave dynamics
as was demonstrated numerically in Ref.~\cite{SVR1993}.
In our case, however, by reducing the model to the FK-ME one,
we can describe the solitary waves analytically and rigorously.

In the simulation
we started from the well-defined initial configuration, when
all contacts are relaxed except the one or two where kink's motion is initiated.
If one starts from a random initial configuration,
we expect that kinks will emerge at random places,
so that several kinks may propagate through the system simultaneously,
as was observed in simulation of the BK model~\cite{SVR1993}.

Also we assumed that all $\lambda_c$-contacts
are characterized by the same $F_s (u)$ dependence and thus
have the same threshold values $F_{\rm th}$.
This is correct if the number of original contacts
within a single $\lambda_c$-contact,
$N_{\lambda} = \left( \lambda_c /a_c \right)^2$, is infinite.
Otherwise, different $\lambda_c$-contacts will have different
threshold values $F_{n}$ however the distribution of their thresholds
is narrower that the distribution of thresholds of single asperities by
a factor $\sqrt{N_{\lambda}}$. A narrow  distribution of thresholds will 
nevertheless have a qualitative effect 
because rupture fronts may stop when they meet
$\lambda_c$-contacts with a threshold above the driving force.
When the interface is disordered,
the avalanches will have finite lengths and may become short for forces
near $f_{\mathrm{ini}}$, for which the rupture fronts propagate at the
minimal velocity.

Our approach may also incorporate the existence of disorder and defects
always present in real materials.
On the one hand, defects may nucleate kinks (cracks);
on the other hand, the kink propagation may be
slowed down up to its complete arrest due to pinning by the defects.
For example, the slowing down of the 1D crack
propagating through a 2D system with
quenched randomly distributed defects was considered in Ref.~\cite{KV2006}.

\smallskip
Thus, reducing the EQ-ME model of friction to the FK-ME one,
we described avalanche-like dynamics of the frictional interface
--- the solitary wave of contacts breaking.
If the force $F_s (u)$ has a sawtooth shape, then
the interface dynamics may be described analytically;
otherwise one has to use numerics.
The analogy with the FK model may be extended even further:
\\
$\bullet$
The driven FK model exhibits hysteresis
when the force increases and then decreases~\cite{BBR1997,BKbook}.
The same effect was observed in
the large-scale crack simulation~\cite{HM1998},
thus could be observed in the frictional interface too.
\\
$\bullet$
Effects of nonzero temperature may be considered.
One may predict that at $T>0$ the sliding kinks will experience
an additional damping, while the immobile (e.g., arrested) kinks
will slowly move (creep) due to thermally activated jumps.
\\
$\bullet$
As shown in Refs.~\cite{BHZ2000,BKbook}, a
fast driven kink begins to oscillate due to excitation of its shape mode,
and then, with the further increase of driving, the kink is destroyed.
This effect is similar to what is observed
in fracture mechanics, where cracks begin to oscillate
and then branch~\cite{FM1999}.
\\
$\bullet$
If the interaction between the atoms is nonlinear and stiff enough,
the FK model admits the existence of
supersonic kinks~\cite{B2000}
which are similar to solitons of the Toda chain.
It would be interesting to study if similar waves may appear
in the frictional interface,
% (although this looks unrealistic),
as was predicted in crack propagation~\cite{GJ2005}.
\\
$\bullet$
One may suppose that the damping coefficient $\eta$
in the equation of motion ~(\ref{crack.eqi00})
depends on the kink velocity, % so that
$\eta (v)$.
In fracture mechanics, this coefficient defines
the rate at which the energy is removed from the crack edge,
thus it plays a crucial role.
\\
$\bullet$
A large number of works is devoted to different
generalizations of the FK model to 2D system (e.g., see~\cite{BKbook}).
For example,
if kinks attract one another in the $y$ (transverse)
direction, they unite into a line (dislocation)
which moves as a whole (or due to secondary kinks).

%=========================================================================
\section{Conclusion}
\label{concl}

We discussed the crucial role in sliding friction
of the elastic interaction between the contacts
at the inhomogeneous frictional interface and proposed various
approaches to treat this problem from different viewpoints.
The interaction produces a characteristic
elastic correlation length $\lambda_c = a^2 E / k_c$.
At distance $r < \lambda_c$ the slider may be considered
as a rigid body but with a strong contacts' interaction,
which leads to shrinking of the effective contact breaking threshold
distribution 
and an enhanced possibility for a mechanical elastic instability to
appear, which 
is conducive to stick slip.
At large distances $r > \lambda_c$, the contact-contact interaction
leads to screening
of local perturbations in the interface, or to appearance
of collective modes (frictional cracks) propagating as solitary waves.

\smallskip
In our work we assumed that the external stress
(the driving force) is uniform across the system.
In a general case, however, stress is nonuniform and may moreover
change with (adjust itself to) interface dynamics,
so that the problem should be considered self-consistently.
For given boundary conditions, determined by the experimental setup,
one should calculate the stress field, e.g., by finite element technique,
which provides the driving force $f (r)$ in the FK-ME model.
The latter defines the displacement field at the interface
through the solution of the FK-ME master equations.
The displacement field in turn is to be used
as the boundary condition for
the elastic-theory equations at the frictional interface
(from other sides of the slider,
the boundary conditions should correspond to a given experimental setup).

%==================================================================
\acknowledgments
We wish to express our gratitude to
E.A.~Jagla, B.N.J.~Persson, M.~Urbakh, and S.~Zapperi
for helpful discussions.
This work was supported in part by
CNRS-Ukraine PICS grant No.~5421,
by ESF Eurocore FANAS AFRI through CNR-Italy,
by PRIN/COFIN 20087NX9Y7,
and by the SNF Sinergia Project NPA1617.
O.B.~acknowledges hospitality at  SISSA and ICTP Trieste.
% ,  and at the University of Milan (Italy).

%==================================================================

\end{document}